\documentclass[%
 reprint,
 amsmath,amssymb,
 aps,
prb,
 longbibliography,
 floatfix,
]{revtex4-2}

\usepackage{graphicx}
\usepackage[dvipsnames,svgnames]{xcolor}
\usepackage{tikz}
\usetikzlibrary{calc,arrows.meta}
\usepackage{physics}

\usepackage{mathtools}
\mathtoolsset{showonlyrefs}

\usepackage{iftex}
\ifPDFTeX
    \usepackage[T1]{fontenc}
    \usepackage[utf8]{inputenc}
    \usepackage{newtxtext}
    \usepackage{newtxmath}
\else
    \usepackage{fontspec}
    \usepackage[math-style=TeX]{unicode-math}
    \IfFileExists{texgyretermes-math.otf}{%
        \setmathfont{texgyretermes-math.otf}%
    }{%
        \IfFileExists{XITSMath-Regular.otf}{%
            \setmathfont{XITSMath-Regular.otf}%
        }{%
            \setmathfont{latinmodern-math.otf}%
        }%
    }%
\fi
\usepackage[colorlinks,linkcolor=blue,citecolor=blue,urlcolor=blue]{hyperref}
\newcommand{\hc}{\hat{c}}
\newcommand{\hH}{\hat{H}}
\newcommand{\eps}{\epsilon}
\newcommand{\hone}{\hat{\mathbb{1}}}
\def\w{\omega}

\newcommand{\RR}{\mathrm{R}}

\newcommand{\nR}{\hat{n}_{\RR}}
\newcommand{\Hsys}{\hat{H}_{\mathrm{DQD}}}
\newcommand{\HQPC}{\hat{H}_{\mathrm{QPC}}}
\newcommand{\HQPCtun}{\hat{H}_{\mathrm{QPC}}^{\mathrm{tun}}}

\begin{document}

\title{Triplet-assisted leakage during singlet-triplet qubit readout with a quantum point contact}

\author{Karol Kawa}
\email{kawa@fzu.cz}
\affiliation{FZU --- Institute of Physics of the Czech Academy of Sciences, Na Slovance 1999/2, 182 00 Prague, Czech Republic}

\begin{abstract}
Quantum point contact readout theory for singlet-triplet qubits in a lateral double quantum dot is extended by including tunneling of triplet configurations into a higher-energy level of the neighboring dot.
This additional channel creates energetically allowed leakage pathways that modify the branch-dependent charge and current-noise signatures, even when the Pauli blockade remains effective within the ground-state manifold.
The model contains two single-particle levels in each dot.
The resulting singlet and triplet block structure is derived together with a Lindblad master equation.
Quantum-jump simulations are then used to resolve the dynamics of individual readout events.
A complementary Liouvillian steady-state analysis identifies the regime in which tunneling to the excited level qualitatively changes the readout signatures, with the crossover determined by the level spacing.
\end{abstract}

\maketitle

\section{Introduction}

Spin qubits in semiconductor quantum dots have developed rapidly since the proposal of Loss and DiVincenzo~\cite{Loss1998}.
In addition to single-spin encoding, a particularly important platform is the singlet-triplet ($S$--$T\,$) qubit~\cite{Levy2002} realized in a lateral double quantum dot (DQD).
Its logical basis is commonly written, in the spin basis of the two dots, as the singlet
$\ket*{S} = \left( \ket*{\uparrow\downarrow} - \ket*{\downarrow\uparrow} \right) / \sqrt{2}$
and the zero-spin-projection triplet
$\ket*{T_0} = \left( \ket*{\uparrow\downarrow} + \ket*{\downarrow\uparrow} \right) / \sqrt{2}$.
This encoding enables control through exchange~\cite{Petta2005} and magnetic field~\cite{Foletti2009} gradients, suppresses sensitivity to spatially uniform magnetic field fluctuations because the logical states share the same spin projection~\cite{Burkard2023}, and can operate at comparatively low magnetic fields, which is attractive for hybrid superconducting and spin-photon architectures~\cite{Burkard2023,Liles2024SingletTripletHoleMOS}.
Implementations span GaAs~\cite{Petta2005,Cerfontaine2020ClosedLoopControl}, Si/SiGe~\cite{Maune2012,Wu2014,Jock2022SiliconSTSpinValley}, and more recently hole-based Si MOS devices~\cite{Liles2024SingletTripletHoleMOS}, while multi-qubit implementations continue to improve in coherence and control fidelity~\cite{Shulman2012EntanglementSTQubits,Burkard2023,Zhang2025UniversalControlFourST}.

The readout of $S$--$T$ qubits is typically based on the Pauli spin blockade (PSB)~\cite{Ono2002}.
In the standard single-level picture, the singlet can access a doubly occupied charge state, whereas the triplet is blocked, so spin information is converted into charge dynamics that can be detected by a nearby electrometer such as a quantum point contact (QPC).
The resulting detector statistics are branch-dependent.
The singlet dynamics allows for charge reorganization, and thus enhances the noise of current flowing through the QPC, while the triplet branch suppresses charge motion and approaches Poissonian detector transport.
Experimentally, QPC charge sensors have resolved time-domain charge dynamics and counting statistics in quantum dots~\cite{Ihn2009ChargeDetection,gustavsson2006_counting_stats_qdot}.
Theoretically, continuous-measurement and backaction approaches have been developed for coupled dots and DQDs monitored by QPCs~\cite{goan2001_cqd_qpc,Stace2004,StaceBarretArxiv,Ouyang2010,Li2013Backaction}.

For spin readout, Barrett and Stace developed a master-equation description of two-spin measurements in exchange-interaction quantum computers~\cite{Barrett_PRB_2006}.
Later, Roszak, Marcinowski, and co-workers included phonon-assisted processes and clarified how relaxation, dephasing, and measurement backaction shape the stochastic current record~\cite{marcinowski_phonon_2013,roszak_decoherence-enhanced_2015}.
However, a common assumption in essentially all of these treatments is that each dot contributes only one active energy level.

That approximation is not always justified.
Realistic dots support a first excited level separated from the ground level by an energy $\eps$ that can be comparable to the QPC bias.
In that regime, a triplet can occupy a doubly occupied configuration provided that one electron tunnels into the excited level, thereby bypassing the usual blockade within the ground-level manifold.
Experimental studies in silicon DQDs have already shown that excited states and phonon-assisted processes can qualitatively modify PSB transport~\cite{Liu2008PauliSB}.
This raises the central question addressed here, namely of how strongly the additional excited level compromises the charge and noise signatures on which $S$--$T$ readout relies.

In this work, this question is addressed with a DQD model containing ground and excited single-particle levels in each dot, coupled to a QPC\@.
The system Hamiltonian is decomposed into singlet and triplet blocks, the corresponding Lindblad dynamics generated by QPC-induced charge fluctuations is derived, and both individual trajectories and steady-state current noise are analyzed.
This extends earlier single-level treatments~\cite{Barrett_PRB_2006,marcinowski_phonon_2013,roszak_decoherence-enhanced_2015} to the experimentally relevant regime in which both ground and excited levels participate.
When the excited level enters the detector window, triplet leakage can strongly enhance the low-frequency noise and even make the triplet branch noisier than the singlet branch, reversing the usual PSB intuition.

The organization of this paper is as follows. 
Section~\ref{sec:model} introduces the model system, including the DQD description with ground and excited levels and its coupling to the QPC\@.
Section~\ref{sec:methods} describes the methods used throughout the work, i.e.\, the Lindblad master equation, the quantum-jump (quantum-trajectory) approach employed to simulate single measurement realizations and the use of the quantum regression theorem to evaluate the QPC current spectrum.
Section~\ref{sec:results} presents the numerical results, focusing on the dependence of the Fano factor on the level spacing $\eps$ and on the conditions under which leakage through the excited level qualitatively alters the spin-dependent noise signatures.
Finally, Sec.~\ref{sec:conclusions} summarizes the main findings and outlines implications for the design of high-fidelity $S$--$T$ readout in realistic quantum-dot qubits with accessible excited levels.

\vfill

\begin{figure*}[t!]
\begin{minipage}{0.47\linewidth}
    \centering
    \includegraphics[width=\linewidth]{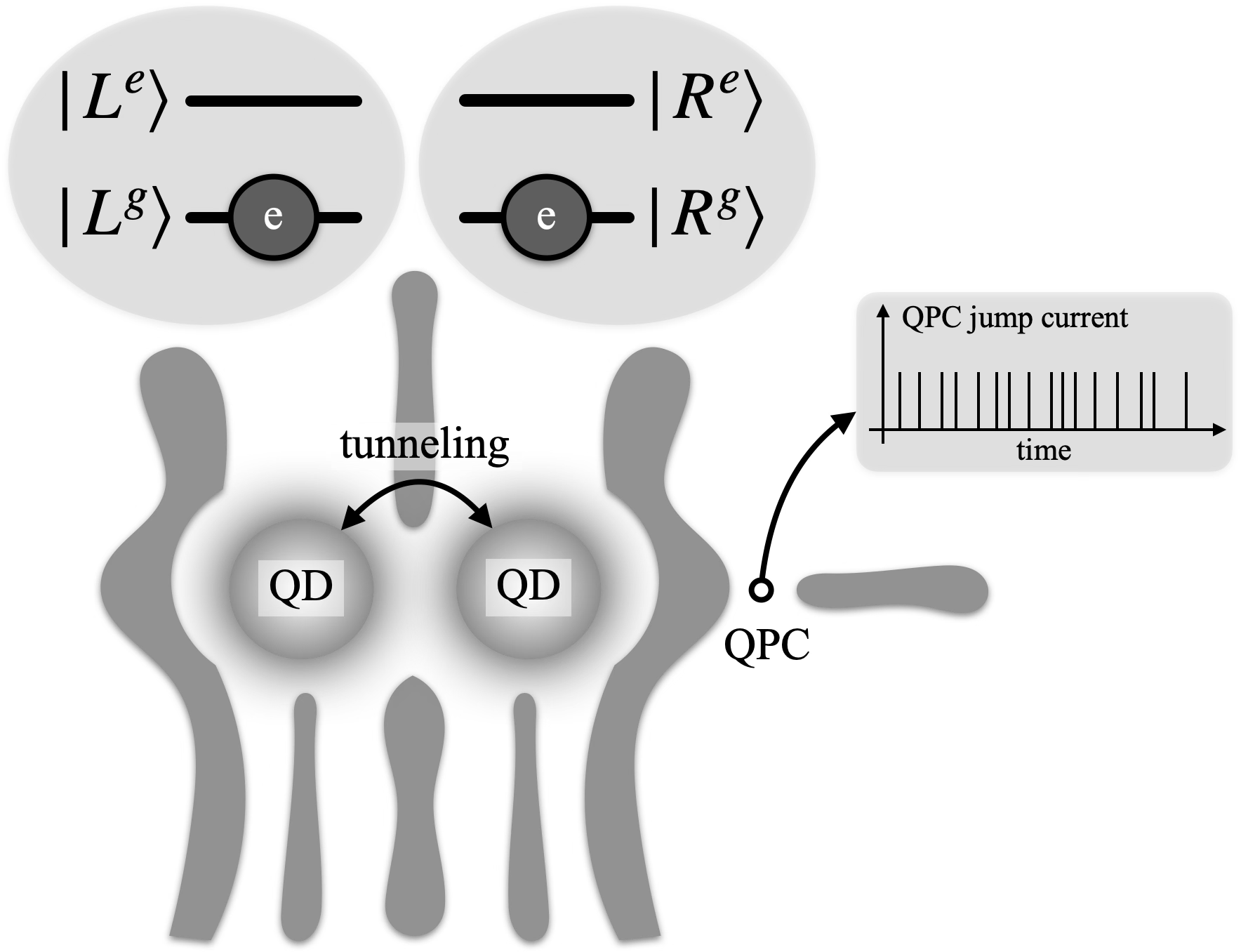}
    \caption{Schematic representation of the lateral double quantum dot capacitively coupled to the QPC detector placed next to the right dot.
    At the top: single particle electron energy levels [cf. Eq.~\eqref{eq:natural-basis}] with an exemplary charge separated configuration in ground manifold.}\label{fig:device-dqd-qpc}
\end{minipage}
\hfill
\begin{minipage}{0.47\linewidth}
    \centering
    \includegraphics[width=0.9\linewidth]{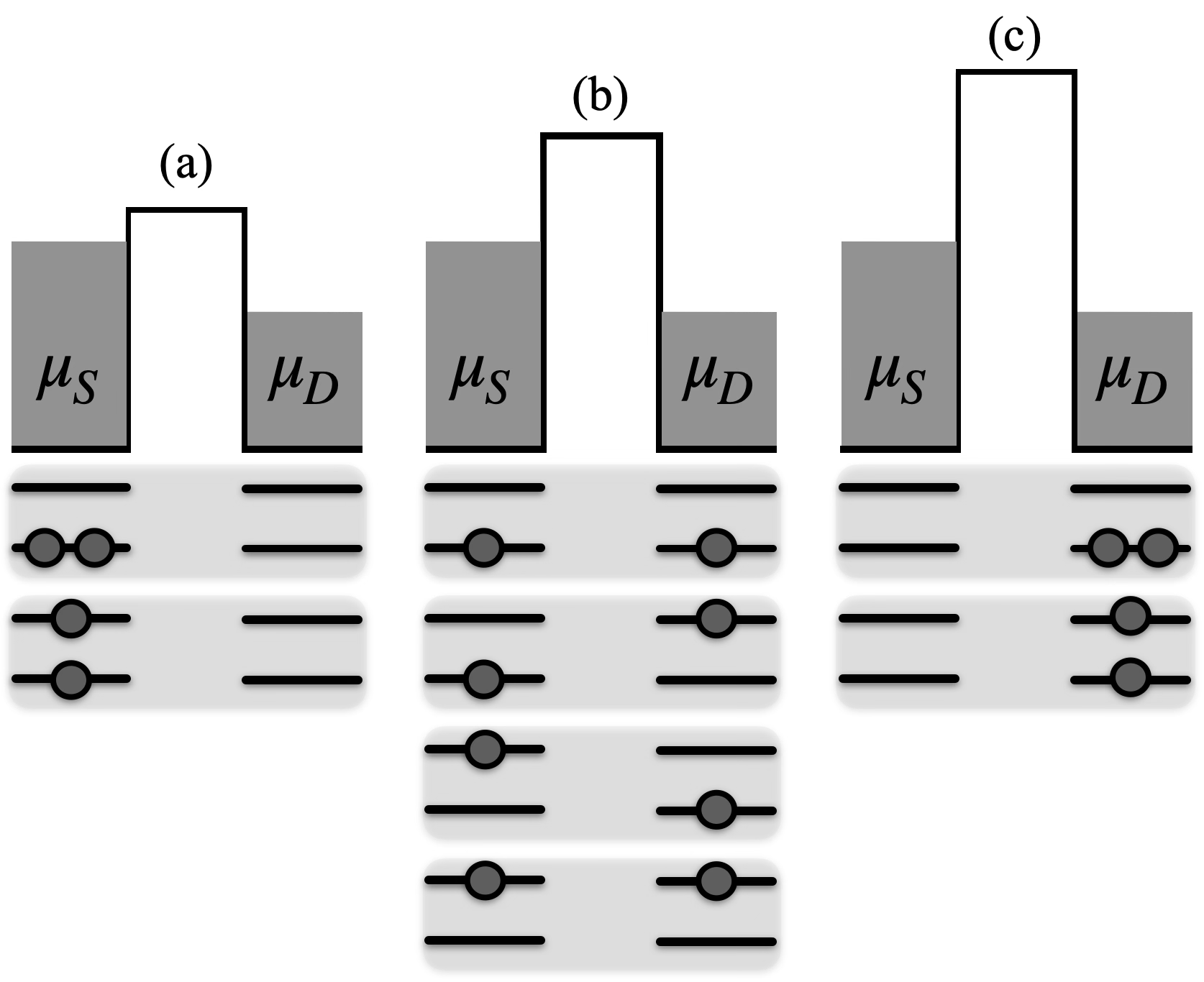}
    \caption{Band diagram illustrating the three relevant DQD charge configurations and the corresponding QPC detector barrier.
    Panels (a)--(c) show, respectively, both electrons localized in the left dot, one electron in each dot, and both electrons localized in the right dot.
    Since the QPC is placed next to the right dot, increasing the right-dot occupation modifies the barrier between the source and drain leads.}\label{fig:device-band-diagram}
\end{minipage}
\end{figure*}

\section{Model system}\label{sec:model}

The model system is described by the total Hamiltonian,
\begin{align}
    \hH = \Hsys + \HQPC + \HQPCtun,
\end{align}
where $\Hsys$ denotes the double quantum dot Hamiltonian, $\HQPC$ the electronic reservoir formed by the source and drain leads of the QPC, and $\HQPCtun$ the QPC tunneling term, including both the bare source-drain transfer and its modulation by the DQD charge.
The following subsections introduce these three contributions, in turn.

\subsection*{Lateral double quantum dot}

Consider a lateral GaAs DQD monitored by a nearby QPC charge sensor, as sketched in Fig.~\ref{fig:device-dqd-qpc}.
Each dot hosts two single-particle energy levels, a ground level ($g$) and a first excited level ($e$).
The corresponding energetic picture, together with the dependence of the QPC barrier on the occupation of the right dot, is illustrated in Fig.~\ref{fig:device-band-diagram}.
The two dots form an artificial molecule with spin-conserving tunneling between them. 
Throughout, the interdot Coulomb repulsion is neglected owing to the relatively large interdot separation, and spin is treated as a passive degree of freedom so that the DQD Hamiltonian is spin independent,
\begin{align}
    \Hsys = \hH_0 + \hH_\mathrm{tun} + \hH_{C}.
    \label{eq:system-hamiltonian}
\end{align}
The first term in Eq.~\eqref{eq:system-hamiltonian},
\begin{align}
    \hH_0 = \eps \sum_{i=L,R} \hat{n}_{i,e},
\end{align}
describes the bare single-particle energies in the two dots, which are taken to be identical.
The reference energy of the ground level in each dot is set to zero, so $\eps$ is the energy of the excited level.
Electrons can tunnel between the dots via spin-conserving processes,
\begin{align}
\begin{aligned}
    \hH_\mathrm{tun} ={} & -\! \sum_{\alpha=g,e} t_{\alpha}
    \left( \hat{c}^{\dagger}_{L\alpha} \hat{c}_{R\alpha} + \mathrm{H.c.} \right) \\
    & - t_c \sum_{\alpha=g,e} \left( \hat{c}_{L\alpha}^\dagger \hat{c}_{R\bar{\alpha}} + \mathrm{H.c.} \right),
\end{aligned}
\end{align}
where $\bar{\alpha}$ denotes the opposite orbital ($\bar{g}=e$, $\bar{e}=g$).
The amplitudes $t_g$ and $t_e$ describe tunneling between ground and excited levels, respectively, while $t_c$ is the cross-level tunneling amplitude between a ground level in one dot and an excited level in the other.
For simplicity, the calculations below set $t_g = t_e = t_c \equiv t$.
Local Coulomb repulsion is given by
\begin{align}
    \hH_{C} = \frac{U}{2} \sum_{i=L,R} \hat{n}_{i} \bigl( \hat{n}_{i} - 1 \bigr),
\label{eq:Hubbard}
\end{align}
with the on-site charging energy penalty $U$ for double occupation of a single dot.
The operators $\hat{c}_{i\alpha}^\dagger$ ($\hat{c}_{i\alpha}$) create (annihilate) an electron in the left ($i=L$) or right ($i=R$) dot and in the ground ($\alpha=g$) or excited ($\alpha=e$) orbital.
The corresponding number operators appearing in the above formulas are given by
\begin{align}
    \hat{n}_{i\alpha} = \hat{c}_{i\alpha}^\dagger \hat{c}_{i\alpha}, \qquad
    \hat{n}_i = \sum_{\alpha} \hat{n}_{i\alpha}.
\end{align}
Because interorbital tunnelings between dots, such as
$\hat{c}_{Lg}^\dagger \hat{c}_{Re}$, are included in $\hH_\mathrm{tun}$, the singlet and triplet symmetry blocks of $\Hsys$ have complex forms taking into account ground, excited, and mixed manifolds.

\subsubsection*{Two-electron states: singlets and triplets}

Since $\Hsys$ conserves the total spin, it decomposes into singlet and triplet blocks with total spin 0 and 1, respectively.
From this point on, kets such as $\ket*{S_0^g}$ and $\ket*{T^g}$ denote the orbital/configurational part of the corresponding antisymmetrized two-electron states; the spin wave function fixes the singlet or triplet symmetry.
The two-particle Hilbert space is spanned by configurational states labeled by the dot index ($L$ or $R$) and orbital ($g$ or $e$),
\begin{align}
\mathcal{H} = \left\{
                 \ket*{i^\alpha\, j^\beta}
              \right\}
            = \left\{
                 \hat{c}_{i\alpha}^\dagger \hat{c}_{j\beta}^\dagger \ket*{\mathrm{vac}}
              \right\},
    \label{eq:natural-basis}
\end{align}
with $i,j\in\{L,R\}$, $\alpha,\beta\in\{g,e\}$, and $\ket*{\mathrm{vac}}$ denoting an empty DQD\@.
With $\Hsys$ [Eq.~\eqref{eq:system-hamiltonian}] and the natural Hilbert space $\mathcal{H}$ [Eq.~\eqref{eq:natural-basis}] as starting points, the block-diagonal Hamiltonian with singlet and triplet spin symmetry is $\Hsys = \hH_\mathrm{singlet} + \hH_\mathrm{triplet}$.

\subsubsection*{Singlet subspace}

First, consider singlets in which both electrons occupy the ground energy level.
The ground state manifold of singlets is given by
\begin{subequations}
\begin{align}
\begin{aligned}
    \ket*{S_-^g} 
        & = \frac{1}{\sqrt{2}} \Bigl( \ket*{L^g L^g} - \ket*{R^g R^g} \Bigr), \\
    \ket*{S_0^g} 
        & = \frac{1}{\sqrt{2}} \Bigl( \ket*{L^g R^g} + \ket*{R^g L^g} \Bigr), \label{eq:qubit-singlet} \\
    \ket*{S_+^g} 
        & = \frac{1}{\sqrt{2}} \Bigl( \ket*{L^g L^g} + \ket*{R^g R^g} \Bigr).
\end{aligned}
\end{align}
\end{subequations}
Among these, $\ket*{S_0^g}$ is the only charge-separated state and belongs to the standard $S$--$T$ qubit basis.
The corresponding singlets belonging to the higher-energy manifold are
\begin{align}
\begin{aligned}
    \ket*{S_-^e} 
        & = \frac{1}{\sqrt{2}} \Bigl( \ket*{L^e L^e} - \ket*{R^e R^e} \Bigr), \\
    \ket*{S_0^e} 
        & = \frac{1}{\sqrt{2}} \Bigl( \ket*{L^e R^e} + \ket*{R^e L^e} \Bigr), \\
    \ket*{S_+^e} 
        & = \frac{1}{\sqrt{2}} \Bigl( \ket*{L^e L^e} + \ket*{R^e R^e} \Bigr).
\end{aligned}
\label{eq:singlets}
\end{align}
Beyond these, there also exist mixed-manifold singlets, that is, with one electron in the ground and the other in the excited energy level (cf. Fig.~\ref{fig:device-band-diagram}),
\begin{align}
\begin{aligned}
    \ket*{2_s^-}  
        & = \frac{1}{2} \Bigl( \ket*{L^g L^e} + \ket*{L^e L^g} - \ket*{R^g R^e} - \ket*{R^e R^g} \Bigr), \\
    \ket*{1_s^-}
        & = \frac{1}{2} \Bigl( \ket*{L^g R^e} + \ket*{R^e L^g} - \ket*{R^g L^e} - \ket*{L^e R^g} \Bigr), \\
    \ket*{2_s^+}  
        & = \frac{1}{2} \Bigl( \ket*{L^g L^e} + \ket*{L^e L^g} + \ket*{R^g R^e}  + \ket*{R^e R^g} \Bigr), \\
    \ket*{1_s^+}
        & = \frac{1}{2} \Bigl( \ket*{L^g R^e} + \ket*{R^e L^g} + \ket*{R^g L^e}  + \ket*{L^e R^g} \Bigr).
\end{aligned}
\end{align}
The singlet part of $\Hsys$ is a block diagonal matrix and decomposes into three independent terms,
\begin{align}
    \hH_\mathrm{singlet} = \hH_\mathrm{singlet}^{(1)} + \hH_\mathrm{singlet}^{(2)} + \hH_\mathrm{singlet}^{(3)}.
\end{align}
The first term is trivial,
\begin{align}
    \hH_\mathrm{singlet}^{(1)} = \left( U + \eps \right) \dyad{s_1},
\end{align}
and consists of the eigenstate $\ket*{s_1} = \ket*{2_s^-}$ with the eigenenergy $E_{s_1} = U + \eps$.

The second singlet term couples ground and excited orbitals,
\begin{align}
    \hH_\mathrm{singlet}^{(2)} = {} & U \dyad*{S_-^g} + (U + 2 \eps) \dyad*{S_-^e} + \eps \dyad*{1_s^-} \\
    & -\sqrt{2} \, t\, \bigl( \dyad*{S_-^g}{1_s^-} + \mathrm{H.c.} \bigr) \\
    & +\sqrt{2} \, t\, \bigl( \dyad*{S_-^e}{1_s^-} + \mathrm{H.c.} \bigr),
\end{align}
and contributes to another three eigenstates,
\begin{align}
    \ket*{s_i} = \sum_{X=S_-^g,S_-^e,1_s^-} P_{iX} \ket*{X}, \quad \text{for} \quad i=2,3,4.
\end{align}
The remaining singlet term is
\begin{align}
\begin{aligned}
   \hH_\mathrm{singlet}^{(3)} ={} & U \dyad*{S_+^g} + 2\eps\dyad*{S_0^e} + (U + 2\eps) \dyad*{S_+^e} \\
   & + \eps\dyad*{1_s^+} + \left( U + \eps \right) \dyad*{2_s^+} \\
   & -2 t\, \bigl( \dyad*{S_0^g}{S_+^g} + \mathrm{H.c.} \bigr)
     - \sqrt{2} t\, \bigl( \dyad*{S_0^g}{2_s^+} + \mathrm{H.c.} \bigr) \\
   & -\sqrt{2}t\, \bigl( \dyad*{S_+^g}{1_s^+} + \mathrm{H.c.} \bigr) \\
   & -2t\, \bigl( \dyad*{S_0^e}{S_+^e} + \mathrm{H.c.} \bigr)
     -\sqrt{2} t\, \bigl( \dyad*{S_+^e}{2_s^+} + \mathrm{H.c.} \bigr) \\
   & - \sqrt{2} t\, \bigl( \dyad*{S_+^e}{1_s^+} + \mathrm{H.c.} \bigr) \\
   & - 2t\, \bigl(\dyad{1_s^+}{2_s^+} + \mathrm{H.c.} \bigr),
\end{aligned}
\end{align}
with eigenstates
\begin{align}
    \ket*{s_i} = \sum_{X} P_{iX} \ket*{X}, \quad i=5,6,\dots,10,
\end{align}
and where $X\in\{ S_0^g, S_0^e, S_+^g, S_+^e, 1_s^+, 2_s^+\}$.
The singlet transition scheme and the corresponding effective jump-rate matrix (see Sec.~\ref{sec:methods} for details) are shown in the upper row of Fig.~\ref{fig:jumps-by-sector}.
Panel (a) shows the complete high-bias graph of off-diagonal QPC-induced transitions between reduced singlet eigenstates; diagonal elastic channels are not drawn.
The singlet states are visually grouped into three manifolds, while the colors of the state lines indicate the three block-Hamiltonian sectors, namely $\{s_1\}$, $\{s_2,s_3,s_4\}$, and $\{s_5,\dots,s_{10}\}$; panel (b) gives the corresponding aggregated jump-rate matrix (cf. Sec.~\ref{sec:methods}).
 
\subsubsection*{Triplet subspace}

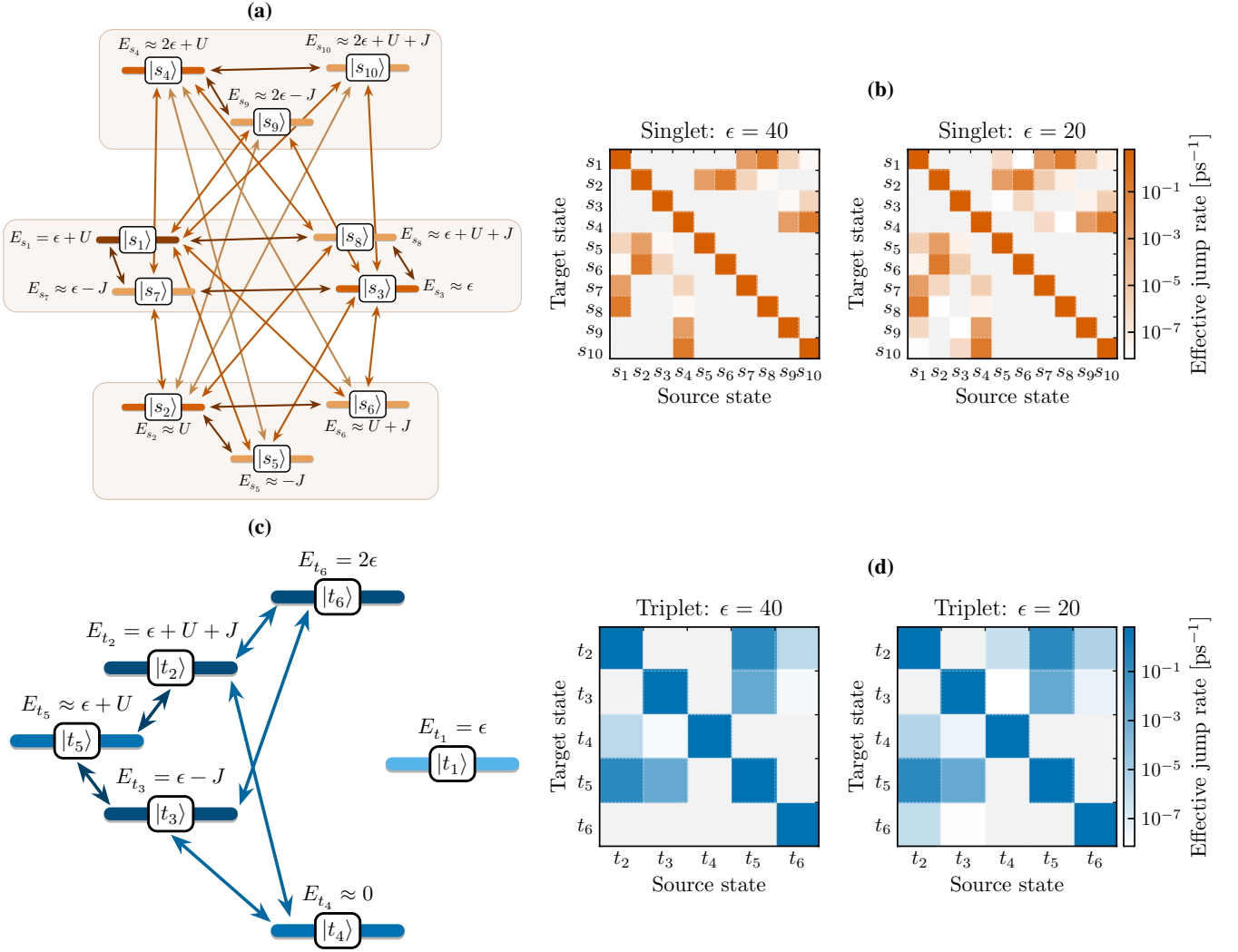
\begin{figure*}[t!]
\begin{minipage}[c]{0.43\linewidth}
   \textbf{(a)}\\[0.25em]
   \centering
   \resizebox{\linewidth}{!}{
\begingroup
\definecolor{singletdark}{HTML}{8C3D00}
\definecolor{singletmid}{HTML}{D55E00}
\definecolor{singletlight}{HTML}{E6A05A}

\begin{tikzpicture}[
  x=0.02cm, y=-0.017cm,
  >=Stealth,
  every node/.style={font=\LARGE},
  levelshadow/.style={draw=black!28, line width=6pt, line cap=round},
  levelB1/.style={draw=singletdark, line width=6pt, line cap=round},
  levelB2/.style={draw=singletmid, line width=6pt, line cap=round},
  levelB3/.style={draw=singletlight, line width=6pt, line cap=round},
  secboxG/.style={draw=singletdark!35, fill=singletdark!5, rounded corners=12pt, line width=0.8pt},
  secboxM/.style={draw=singletdark!35, fill=singletdark!5, rounded corners=12pt, line width=0.8pt},
  secboxE/.style={draw=singletdark!35, fill=singletdark!5, rounded corners=12pt, line width=0.8pt},
  arrowclearance/.style={shorten <=2pt, shorten >=2pt},
  intramanifold/.style={draw=singletdark!85!black, line width=1.7pt, arrowclearance},
  adjacentmanifold/.style={draw=singletmid!90!black, line width=1.7pt, arrowclearance},
  remotemanifold/.style={draw=singletlight!85!black, line width=1.7pt, arrowclearance},
]

\def\statehalflen{60}

\def\excitedManifoldY{25}
\def\mixedManifoldY{337}
\def\groundManifoldY{645}
\def\energyJStep{5}
\def\energyUStep{90}
\def\energyUPlusJStep{95}

\path[secboxE] (120, \excitedManifoldY-75)  rectangle (655, \excitedManifoldY+145);
\path[secboxM] (-30, \mixedManifoldY-37)    rectangle (780, \mixedManifoldY+133);
\path[secboxG] (110, \groundManifoldY-45)   rectangle (650, \groundManifoldY+170);

\coordinate (s4C)  at (220, \excitedManifoldY);
\coordinate (s10C) at (540, \excitedManifoldY-\energyJStep);
\coordinate (s9C)  at (390, \excitedManifoldY+\energyUPlusJStep);

\coordinate (s1C)  at (180, \mixedManifoldY);
\coordinate (s8C)  at (520, \mixedManifoldY-\energyJStep);
\coordinate (s7C)  at (205, \mixedManifoldY+\energyUPlusJStep);
\coordinate (s3C)  at (555, \mixedManifoldY+\energyUStep);

\coordinate (s2C)  at (220, \groundManifoldY);
\coordinate (s6C)  at (540, \groundManifoldY-\energyJStep);
\coordinate (s5C)  at (390, \groundManifoldY+\energyUPlusJStep);


\foreach \name in {s1,s2,s3,s4,s5,s6,s7,s8,s9,s10}{
  \coordinate (\name L) at ($(\name C)+(-\statehalflen,0)$);
  \coordinate (\name R) at ($(\name C)+(\statehalflen,0)$);
}

\draw[remotemanifold,<->] ($(s4C)+(9,28)$)    -- ($(s5C)+(-10,-29)$);     
\draw[remotemanifold,<->] ($(s4C)+(28,25)$)   -- ($(s6C)+(-24,-27)$);     
\draw[remotemanifold,<->] ($(s2C)+(19,-28)$)  -- ($(s9C)+(-20,27)$);      
\draw[remotemanifold,<->] ($(s2C)+(39,-18)$)  -- ($(s10C)+(-25,30)$);     

\draw[adjacentmanifold,<->] ($(s1C)+(50,-8)$)   -- ($(s9C)+(-33,11)$);   
\draw[adjacentmanifold,<->] ($(s4C)+(-10,27)$)  -- ($(s7C)+(0,-27)$);    
\draw[adjacentmanifold,<->] ($(s3C)+(-0,-27)$) -- ($(s10C)+(0,27)$);  
\draw[adjacentmanifold,<->] ($(s4C)+(47,22)$)   -- ($(s8C)+(-63,-8)$);   
\draw[adjacentmanifold,<->] ($(s3C)+(-26,-25)$) -- ($(s9C)+(26,25)$);    
\draw[adjacentmanifold,<->] ($(s1C)+(68,0)$)    -- ($(s10C)+(-39,22)$);  

\draw[adjacentmanifold,<->] ($(s1C)+(55,8)$)   -- ($(s5C)+(-33,-10)$);  
\draw[adjacentmanifold,<->] ($(s1C)+(68,5)$)   -- ($(s6C)+(-33,-10)$);  
\draw[adjacentmanifold,<->] ($(s3C)+(-33,10)$) -- ($(s5C)+(0,-27)$);   
\draw[adjacentmanifold,<->] ($(s3C)+(0,27)$)   -- ($(s6C)+(0,-27)$);    
\draw[adjacentmanifold,<->] ($(s2C)+(0,-27)$)  -- ($(s7C)+(0,27)$);     
\draw[adjacentmanifold,<->] ($(s2C)+(58,-15)$) -- ($(s8C)+(-33,11)$);   

\draw[intramanifold,<->] ($(s2C)+(60,10)$) -- ($(s5C)+(-60,-7)$);      
\draw[intramanifold,<->] ($(s2C)+(69,0)$)  -- ($(s6C)+(-69,0)$);        

\draw[intramanifold,<->] ($(s1C)+(-45,8)$) -- ($(s7C)+(-45,-8)$);       
\draw[intramanifold,<->] ($(s1C)+(78,2)$)  -- ($(s8C)+(-70,0)$);        
\draw[intramanifold,<->] ($(s3C)+(-69,0)$) -- ($(s7C)+(69,0)$);         
\draw[intramanifold,<->] ($(s3C)+(60,-12)$) -- ($(s8C)+(57,12)$);     

\draw[intramanifold,<->] ($(s4C)+(65,8)$) -- ($(s9C)+(-65,-8)$);        
\draw[intramanifold,<->] ($(s4C)+(70,0)$) -- ($(s10C)+(-70,0)$);          

\node[font=\Large] at ($(s2C)+(0,45)$) {$E_{s_2} \approx U$};
\node[font=\Large] at ($(s6C)+(0,45)$) {$E_{s_6} \approx U+J$};
\node[font=\Large] at ($(s5C)+(0,45)$) {$E_{s_5} \approx -J$};

\node[font=\Large, anchor=east] at ($(s1C)+(-\statehalflen-5,0)$) {$E_{s_1} = \epsilon+U$};
\node[font=\Large, anchor=east] at ($(s7C)+(-\statehalflen-5,0)$) {$E_{s_7} \approx \epsilon - J$};
\node[font=\Large, anchor=west] at ($(s8C)+(\statehalflen+5,0)$) {$E_{s_8}\approx \epsilon+U+J$};
\node[font=\Large, anchor=west] at ($(s3C)+(\statehalflen+5,0)$) {$E_{s_3} \approx \epsilon$};

\node[font=\Large] at ($(s4C)+(0, -44)$) {$E_{s_4} \approx 2\epsilon+U$};
\node[font=\Large] at ($(s10C)+(0,-44)$) {$E_{s_{10}} \approx 2\epsilon+U+J$};
\node[font=\Large] at ($(s9C)+(0, -44)$) {$E_{s_9} \approx 2\epsilon - J$};

\foreach \L/\R/\sty in {
  s1L/s1R/levelB1,%
  s2L/s2R/levelB2,%
  s3L/s3R/levelB2,%
  s4L/s4R/levelB2,%
  s5L/s5R/levelB3,%
  s6L/s6R/levelB3,%
  s7L/s7R/levelB3,%
  s8L/s8R/levelB3,%
  s9L/s9R/levelB3,%
  s10L/s10R/levelB3%
}{
  \draw[levelshadow] ($(\L)+(0,2)$) -- ($(\R)+(0,2)$);
  \draw[{\sty}] (\L) -- (\R);
}

\node[draw, very thick, fill=white, anchor=center, rounded corners] at ($(s2C)$) {$|s_2\rangle$};
\node[draw, very thick, fill=white, anchor=center, rounded corners] at ($(s6C)$) {$|s_6\rangle$};
\node[draw, very thick, fill=white, anchor=center, rounded corners] at ($(s5C)$) {$|s_5\rangle$};

\node[draw, very thick, fill=white, anchor=center, rounded corners] at ($(s1C)$) {$|s_1\rangle$};
\node[draw, very thick, fill=white, anchor=center, rounded corners] at ($(s8C)$) {$|s_8\rangle$};
\node[draw, very thick, fill=white, anchor=center, rounded corners] at ($(s7C)$) {$|s_7\rangle$};
\node[draw, very thick, fill=white, anchor=center, rounded corners] at ($(s3C)$) {$|s_3\rangle$};

\node[draw, very thick, fill=white, anchor=center, rounded corners] at ($(s4C)$) {$|s_4\rangle$};
\node[draw, very thick, fill=white, anchor=center, rounded corners] at ($(s10C)$) {$|s_{10}\rangle$};
\node[draw, very thick, fill=white, anchor=center, rounded corners] at ($(s9C)$) {$|s_9\rangle$};

\end{tikzpicture}
\endgroup}
\end{minipage}\hfill
\begin{minipage}[c]{0.56\linewidth}
    \textbf{(b)}\\[0.25em]
    \centering
    \includegraphics[width=\linewidth]{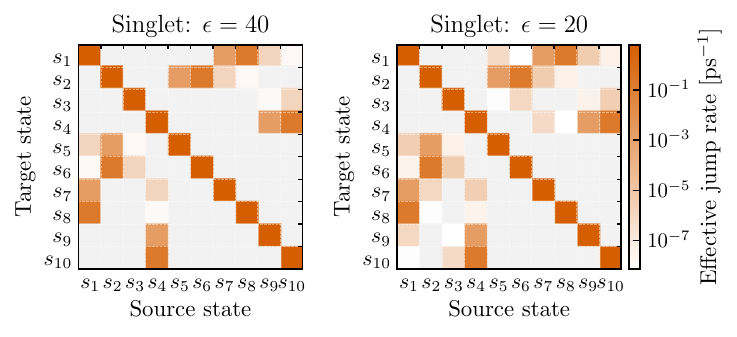}
\end{minipage}

\vspace{0.75em}

\begin{minipage}[c]{0.43\linewidth}
   \textbf{(c)}\\[0.25em]
   \centering
   \resizebox{\linewidth}{!}{\begingroup
\definecolor{tripletdark}{HTML}{004C7A}
\definecolor{tripletmid}{HTML}{0072B2}
\definecolor{tripletlight}{HTML}{56B4E9}

\begin{tikzpicture}[
  x=0.68cm, y=0.68cm,
  >=Stealth,
  every node/.style={font=\normalsize},
  levelshadow/.style={draw=black!30, line width=6pt, line cap=round},
  levelA/.style={draw=tripletdark, line width=6pt, line cap=round},
  levelB/.style={draw=tripletmid, line width=6pt, line cap=round},
  levelC/.style={draw=tripletlight, line width=6pt, line cap=round},
  trans/.style={draw=black, line width=0.9pt},
  intramanifold/.style={draw=tripletdark!85!black, line width=1.5pt},
  adjacentmanifold/.style={draw=tripletmid!90!black, line width=1.5pt}
]

\def\statehalflen{1.45}

\coordinate (t5C) at (-6,  0.75-0.2);
\coordinate (t3C) at (-3.75, -1.0-0.2);
\coordinate (t2C) at (-3.75,  2.5-0.2);

\coordinate (t6C) at ( 0.25,  4.0);
\coordinate (t4C) at ( 0.25, -4.0);

\coordinate (t1C) at ( 3., 0.0);

\foreach \name in {t1,t2,t3,t4,t5,t6}{
  \coordinate (\name L) at ($(\name C)+(-\statehalflen,0)$);
  \coordinate (\name R) at ($(\name C)+(\statehalflen,0)$);
}

\draw[levelshadow] ($(t6L)+(0,-0.06)$) -- ($(t6R)+(0,-0.06)$);
\draw[levelA] (t6L) -- (t6R);
\draw[levelshadow] ($(t3L)+(0,-0.06)$) -- ($(t3R)+(0,-0.06)$);
\draw[levelA] (t3L) -- (t3R);
\draw[levelshadow] ($(t2L)+(0,-0.06)$) -- ($(t2R)+(0,-0.06)$);
\draw[levelA] (t2L) -- (t2R);
\draw[levelshadow] ($(t5L)+(0,-0.06)$) -- ($(t5R)+(0,-0.06)$);
\draw[levelB] (t5L) -- (t5R);
\draw[levelshadow] ($(t4L)+(0,-0.06)$) -- ($(t4R)+(0,-0.06)$);
\draw[levelB] (t4L) -- (t4R);
\draw[levelshadow] ($(t1L)+(0,-0.06)$) -- ($(t1R)+(0,-0.06)$);
\draw[levelC] (t1L) -- (t1R);

\draw[intramanifold,<->] ($(t5R)+(0.,0.2)$) -- ($(t2C)+(0.,-0.5)$);
\draw[intramanifold,<->] ($(t5C)+(0.0,-0.5)$) -- ($(t3L)+(0.,0.2)$);

\draw[adjacentmanifold,<->] ($(t2R)+(0,-0.3)$) -- ($(t4L)+(0.31,0.32)$);
\draw[adjacentmanifold,<->] ($(t4L)+(-0.15,0.2)$) -- ($(t3C)+(0., -0.5)$);
\draw[adjacentmanifold,<->] ($(t2R)+(0.1,0.2)$) -- ($(t6L)+(0.0,-0.2)$);
\draw[adjacentmanifold,<->] ($(t3R)+(0.2,0.2)$) -- ($(t6C)+(-0.7,-0.3)$);

\node[draw, very thick, anchor=center, fill=white, rounded corners] at (t2C) {$|t_2\rangle$};
\node[font=\normalsize, anchor=center] at ($(t2C)+(-0.2, 0.8)$) {$E_{t_2}=\epsilon+U+J$};
\node[draw, very thick, anchor=center, fill=white, rounded corners] at (t6C) {$|t_6\rangle$};
\node[font=\normalsize, anchor=center] at ($(t6C)+(0, 0.8)$) {$E_{t_6} = 2\epsilon$};
\node[draw, very thick, anchor=center, fill=white, rounded corners] at (t3C) {$|t_3\rangle$};
\node[font=\normalsize, anchor=center] at ($(t3C)+(0,0.8)$) {$E_{t_3} = \epsilon-J$};
\node[font=\normalsize, anchor=center] at ($(t5C)+(0.0,0.8)$) {$E_{t_5}\approx \epsilon + U$};
\node[draw, very thick, fill=white, anchor=center, rounded corners]  at (t5C) {$|t_5\rangle$};
\node[font=\normalsize, anchor=center] at ($(t4C)+(0,0.8)$) {$E_{t_4}\approx 0$};
\node[draw, very thick, anchor=center, fill=white, rounded corners]  at (t4C) {$|t_4\rangle$};
\node[draw, very thick, anchor=center, fill=white, rounded corners] at (t1C) {$|t_1\rangle$};
\node[font=\normalsize,anchor=center] at ($(t1C)+(0,0.8)$) {$E_{t_1}=\epsilon$};

\end{tikzpicture}
\endgroup}
\end{minipage}\hfill
\begin{minipage}[c]{0.56\linewidth}
    \textbf{(d)}\\[0.25em]
    \centering
    \includegraphics[width=\linewidth]{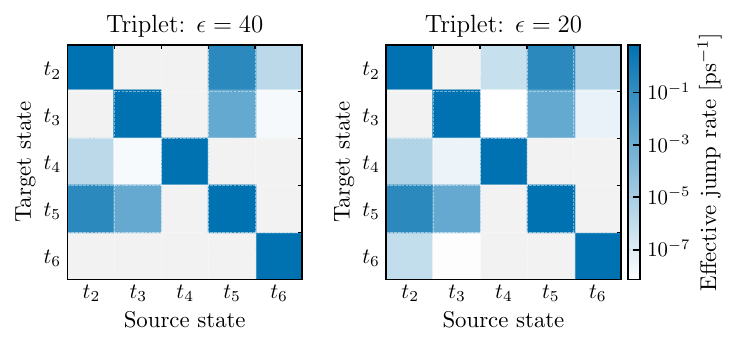}
\end{minipage}
\caption{
  QPC-induced transitions and aggregated jump-rate matrices in the reduced symmetry sectors.
  Panel (a) shows the complete high-bias singlet graph of off-diagonal QPC-induced transitions between reduced singlet eigenstates; panel (c) shows schematic QPC-induced pathways in the triplet sector; and panels (b) and (d) show the corresponding aggregated jump-rate matrices for $\eps=40$~meV and $\eps=20$~meV on a common color scale within each symmetry sector.
  In the schematic panels, darker arrow saturations denote transitions within the same manifold, intermediate saturations denote transitions between adjacent manifolds, and the lightest saturation denotes transitions between remote manifolds; this color coding also reflects the typical strength hierarchy of the corresponding transitions.
  In panel (a), the singlet eigenstates are grouped visually into three manifolds, while the colors of the state lines mark the three block-Hamiltonian sectors $\{s_1\}$, $\{s_2,s_3,s_4\}$, and $\{s_5,\dots,s_{10}\}$; the approximate (or exact if possible) energies are indicated next to the levels.
  In panel (c), the triplet schematic shows the full six-state block structure, whereas panel (d) shows the reduced five-state sector $\{t_2,\dots,t_6\}$ used in the steady-state analysis.}\label{fig:jumps-by-sector}
\end{figure*}

Next, consider triplets.
There is one ground-level triplet,
\begin{align}
    \ket*{T^g} 
        & = \frac{1}{\sqrt{2}} \Bigl( \ket*{L^g R^g} - \ket*{R^g L^g} \Bigr).
\end{align}
There are also four triplets with one electron occupying the ground level and the other occupying the excited level,
\begin{align}
\begin{aligned}
    \ket*{1_t^-} 
        & = \frac{1}{2} \Bigl( \ket*{L^g R^e} - \ket*{R^e L^g} - \ket*{R^g L^e}  + \ket*{L^e R^g} \Bigr), \\
    \ket*{2_t^+} 
        & = \frac{1}{2} \Bigl( \ket*{L^g L^e} - \ket*{L^e L^g} +  \ket*{R^g R^e} - \ket*{R^e R^g} \Bigr), \\
    \ket*{1_t^+} 
        & = \frac{1}{2} \Bigl( \ket*{L^g R^e} - \ket*{R^e L^g} + \ket*{R^g L^e}  - \ket*{L^e R^g} \Bigr), \\
    \ket*{2_t^-} 
        & = \frac{1}{2} \Bigl( \ket*{L^g L^e} - \ket*{L^e L^g} - \ket*{R^g R^e} + \ket*{R^e R^g} \Bigr).
\end{aligned}
\label{eq:triplet-mixed-manifold}
\end{align}
Finally, there is one triplet in the excited-level manifold,
\begin{align}
    \ket*{T^e} 
        & = \frac{1}{\sqrt{2}} \Bigl( \ket*{L^e R^e} - \ket*{R^e L^e} \Bigr).
\end{align}
In the triplet sector, $\Hsys$ decomposes into three independent terms,
\begin{align}
    \hH_\mathrm{triplet} = \hH_\mathrm{triplet}^{(1)} + \hH_\mathrm{triplet}^{(2)} + \hH_\mathrm{triplet}^{(3)}.
\end{align}
The first term is trivial,
\begin{align}
    \hH_\mathrm{triplet}^{(1)} = \eps \dyad*{t_1},
\end{align}
with eigenstate $\ket*{t_1} = \ket*{1_t^-}$ and eigenenergy equal to the excited-level energy $E_{t_1} = \eps$.
The second term is effectively $2\times2$ and can be written as
\begin{align}
    \hH_\mathrm{triplet}^{(2)} = \left( \eps + J + U \right) \dyad*{t_2}
    + \left(\eps - J\right) \dyad*{t_3},
\end{align}
where the eigenstates are
\begin{align}
  \begin{aligned}
    \ket*{t_2} & = \cos\frac{\theta}{2} \ket*{2_t^+} - \sin\frac{\theta}{2} \ket*{1_t^+}, \quad E_{t_2} = \eps + J + U \\
    \ket*{t_3} & = \sin\frac{\theta}{2} \ket*{2_t^+} + \cos\frac{\theta}{2} \ket*{1_t^+}, \quad E_{t_3} = \eps - J,
  \end{aligned}
\end{align}
with $\theta = \arctan\!\left(4 t/U \right)$.
Here, $J = (\sqrt{U^2 + 16 t^2} - U)/2$ denotes an effective exchange contribution originating from virtual tunneling processes.
The remaining triplet states form a $3\times3$ block,
\begin{align}
  \begin{aligned}
    \hH_\mathrm{triplet}^{(3)} ={} &  2 \eps \dyad{T^e} + (U + \eps) \dyad{2_t^-} \\
    & - \sqrt{2}\, t \left( \dyad*{T^g}{2_t^-} + \mathrm{H.c.} \right) \\
    & + \sqrt{2}\, t \left( \dyad*{T^e}{2_t^-} + \mathrm{H.c.} \right),
  \end{aligned}
\end{align}
which is diagonalized as
\begin{align}
    \ket*{t_i} = \sum_{X=T^g,T^e,2_t^-} Q_{iX} \ket*{X}, \quad i=4,5,6.
\end{align}
Again, the corresponding triplet transition scheme and the effective QPC-induced jump-rate matrix (see Sec.~\ref{sec:methods}) are shown in the lower row of Fig.~\ref{fig:jumps-by-sector}.
Panel (c) summarizes representative pathways in the full six-state triplet block, while panel (d) shows the reduced five-state sector $\{t_2,\dots,t_6\}$ used in the steady-state spectral analysis below.

\subsection*{Quantum point contact}

The DQD charge configuration is monitored by a nearby QPC, which acts as a mesoscopic charge sensor.
The QPC is modeled as two noninteracting electron reservoirs, i.e., an electron source ($S$) and an electron drain ($D$) (cf. Fig.~\ref{fig:device-dqd-qpc}).
The bare QPC Hamiltonian reads,
\begin{align}
    \HQPC = \sum_{\ell\in\{S,D\},k} \hbar\w_{\ell k}\,
    \hat{d}^{\dagger}_{\ell k}\hat{d}_{\ell k},
\end{align}
where $\hat{d}^{\dagger}_{\ell k}$ creates an electron with wave number $k$ in lead $\ell\in\{S,D\}$ (the spin index is again suppressed).
The QPC current is sensitive to the right-dot occupation.
Accordingly, the source-drain tunneling through the QPC is described by
\begin{align}
    \HQPCtun = \sum_{kq} \bigl(T_{kq} + \chi_{kq}\,\nR\bigr)
    \bigl(\hat{d}^{\dagger}_{S k}\hat{d}_{D q} + \mathrm{H.c.}\bigr),
\end{align}
where $T_{kq}=T_{qk}$ is the bare tunneling amplitude through the QPC and $\chi_{kq}\,\nR$ is the occupation-dependent correction induced by the DQD-QPC coupling~\cite{Barrett_PRB_2006}. 
Thus, $\HQPCtun$ describes transport through the QPC, with a tunneling amplitude modulated by the charge in the right dot.
In the eigenbasis of $\Hsys$, the operator $\nR$ contains both diagonal matrix elements (generating pure dephasing due to QPC shot noise) and off-diagonal matrix elements that mediate inelastic transitions between DQD eigenstates.

\section{Methods}\label{sec:methods}

\subsection*{Lindblad master equation}

Under a standard Born–Markov–secular treatment, the reduced density matrix of DQD $\rho$ obeys
\begin{equation}
    \dot{\rho} = -(i/\hbar) [\Hsys,\rho] + \sum_{\alpha} \mathcal{D}[\hat{c}_{\alpha}]\,\rho \equiv \mathcal{L} \rho
    \label{eq:lindblad-master-equation}
\end{equation}
where $\mathcal{L}$ is the Liouvillian superoperator, $\mathcal{D}[\hat{o}]\,\rho = \mathcal{J}[\hat{o}]\rho - \frac{1}{2}\{\hat{o}^{\dagger}\hat{o},\rho\}$,
$\mathcal{J}[\hat{o}]\rho = \hat{o} \rho \hat{o}^\dagger$,
and
$\alpha$
labels the retained QPC jump channel.
The jump operators are numerous, so only their general form is listed, and it follows from Fermi's golden rule.
Following Ref.~\cite{Barrett_PRB_2006}, the QPC densities of states are absorbed into the effective tunneling amplitudes $\mathcal{T} = \sqrt{4\pi g_S g_D}\,T_{\mathrm{av}}$ and $\nu = \sqrt{4\pi g_S g_D}\,\chi_{\mathrm{av}}$, where $g_S$ and $g_D$ are the source and drain densities of states at the Fermi energy, while $T_{kq}\approx T_{\mathrm{av}}$ and $\chi_{kq}\approx \chi_{\mathrm{av}}$ are assumed to slowly vary throughout the transport window.
For an inelastic transition-resolved channel $\alpha=(f,i,\eta)$, with final DQD eigenstate $f$, initial DQD eigenstate $i$, and QPC tunneling direction $\eta=\pm$, the jump operators read
\begin{align}
    \hat{c}_{fi,\eta} = \sqrt{\Gamma_\eta(\w_{fi})}\, \mel*{f}{\mathcal{T} + \nu \hat{n}_R}{i} \dyad*{f}{i},
    \label{eq:jump-operators}
\end{align}
where $\hbar\w_{fi} = E_f - E_i$ is the energy difference between the final and initial DQD eigenstates, $V = (\mu_S - \mu_D)$ is the QPC bias energy,
\begin{align}
    \Gamma_\eta(\w) = H(\eta V/\hbar - \w), \quad \eta=\pm,
\end{align}
and
\begin{align}
    H(x) = \left( x + \abs{x} \right) / 2
\end{align}
is the ramp function that encodes the energetically allowed tunneling window.
Here, $\eta=+$ labels the forward QPC tunneling direction $S\to D$, while $\eta=-$ labels the backward direction $D\to S$.
Thus, for each resolved transition $f\leftarrow i$, both forward and backward QPC channels are retained whenever they are energetically allowed.
The calculations reported here keep the inelastic channels transition-resolved: the channel label is the transition pair together with the QPC direction, and transitions are not grouped by numerical proximity of their Bohr frequencies.
For the reduced models used below, no exact nonzero Bohr degeneracy among distinct transition pairs occurs, so every retained inelastic jump operator contains a single off-diagonal projector $\dyad*{f}{i}$.
The zero-Bohr-frequency elastic contribution is the exception retained in grouped form: all diagonal projectors are collected into a single forward elastic jump operator, corresponding to $\eta=+$ and $\w_{fi}=0$; the backward elastic channel vanishes for the positive QPC bias used here.
The inelastic transitions in the singlet and triplet subspaces are summarized in Fig.~\ref{fig:jumps-by-sector}.
Downward transitions are always open, while upward transitions require a QPC bias large enough to supply the relevant eigenenergy difference $V/\hbar > \w_{fi}$ for $\w_{fi}>0$.
The inclusion of mixed and excited triplet configurations therefore creates additional inelastic pathways that are absent in the standard single-level description.
In the parameter range studied below, the relevant transition energies $\w_{fi}$ remain well separated from the QPC-induced rates, so the secular treatment is used consistently.
The very slow structures that appear in the spectra arise from weak effective couplings between manifolds inside the resulting Markovian Liouvillian.
This mechanism underlies the super-Poissonian triplet noise discussed in Sec.~\ref{sec:results}.

\subsection*{Quantum jump method}

The quantum-jump method (Monte Carlo wave-function method)~\cite{Dalibard1992,Carmichael1993} is used to simulate single measurement realizations.
It unravels the Lindblad master equation~\eqref{eq:lindblad-master-equation} into stochastic trajectories of vector states $\ket*{\psi(t)}$.

The evolution alternates between deterministic non-unitary propagation and stochastic jumps associated with the monitored QPC channels.
The continuous part is generated by the effective non-Hermitian Hamiltonian
\begin{align}
    \hH_{\text{eff}} = \Hsys - \frac{i\hbar}{2} \sum_{\alpha} \hc_{\alpha}^\dagger \hc_{\alpha}.
    \label{eq:effective-hamiltonian}
\end{align}
In a short time step $dt$, the no-jump propagation first produces the unnormalized state
\begin{align}
    \ket*{\tilde{\psi}(t+dt)} = \left(1 - \frac{i}{\hbar} \hH_{\text{eff}} dt\right) \ket*{\psi(t)}.
\end{align}
Since $\hH_{\text{eff}}$ is non-Hermitian, the norm decreases by the total jump
probability
\begin{align}
    dp = \sum_{\alpha} dp_{\alpha}, \quad
    dp_{\alpha} = dt\, \mel*{\psi(t)}{\hc_{\alpha}^\dagger \hc_{\alpha}}{\psi(t)}.
\end{align}
Here, $dp_{\alpha}$ is the probability that a jump occurs in channel $\alpha$ during the interval $dt$.
If no jump occurs during the interval $dt$, the state is first propagated over the full step with $\hH_{\text{eff}}$, and the conditional no-jump state is obtained only afterward by normalizing the result,
\begin{align}
    \ket*{\psi(t+dt)} =
    \frac{\ket*{\tilde{\psi}(t+dt)}}{\sqrt{1-dp}}.
\end{align}
If a jump does occur, the channel $\alpha$ is selected according to the relative
weights $dp_{\alpha}/dp$ and the state is updated as
\begin{align}
    \ket*{\psi(t+dt)} =
    \frac{\hc_{\alpha} \ket*{\psi(t)}}{\sqrt{\mel{\psi(t)}{\hc_{\alpha}^\dagger \hc_{\alpha}}{\psi(t)}}}.
\end{align}
The sequence of such stochastic updates generates a measurement record in which
each accepted jump corresponds to a forward or backward QPC tunneling event.
Ensemble averages over many trajectories reproduce the Lindblad dynamics, while
single trajectories provide the time-domain current traces analyzed in Sec.~\ref{sec:results}.

\subsection*{Steady state and the current power spectrum}

The statistics of the current through the QPC are characterized by the two-time current-current correlation function
\begin{equation}
    G(\tau)
    = \big\langle I(t+\tau)\, I(t) \big\rangle
      - \big\langle I(t+\tau)\big\rangle \big\langle I(t)\big\rangle ,
\end{equation}
which, in the stationary regime, depends only on the time difference $\tau$.
Using the two-sided counting-current convention employed in the calculations below, the corresponding current-noise spectrum is
\begin{equation}
    S(\w) = \int_{-\infty}^{\infty} \! d\tau \, e^{-i\w\tau}\, G(\tau).
\end{equation}

Within the quantum-jump description of the detector, tunneling events through the QPC are described by a set of transition-resolved jump operators $\hc_\alpha$~\eqref{eq:jump-operators}.
Thus, it is convenient to introduce the QPC current superoperator
\begin{equation}
    \hat{I}\rho = e \sum_{\alpha} \mathcal{J}[\hc_\alpha]\rho,
\end{equation}
where $e$ is the elementary charge.
The measured (classical) detector current is then the expectation value
\begin{equation}
    I(t) = \Tr{\hat{I}\rho(t)}.
\end{equation}
The corresponding stationary average current is
\begin{equation}
    \overline{I} = \Tr{\hat{I} \rho_\infty },
\end{equation}
where $\rho_\infty$ is the steady-state density matrix which, by definition, obeys $\mathcal{L}\rho_\infty = 0$.

The correlation function $G(\tau)$ is evaluated with the quantum regression theorem (QRT)~\cite{lax1963_formal,breuer2002_open_quantum_systems,gardiner2000_quantum_noise,carmichael1999_stat_methods1}, which
provides a direct connection between single-time dynamics and multi-time correlation
functions in Markovian open quantum systems.
For a reduced density matrix $\rho(t)$ evolving as $\dot{\rho}(t) = \mathcal{L}\rho(t)$,
the QRT states that for arbitrary system operators $A$ and $B$,
\begin{equation}
    \langle A(t+\tau) B(t)\rangle
    = \Tr{%
        A\, e^{\mathcal{L}\tau}\!\left[ B \rho(t) \right]
      },
    \qquad \tau \ge 0,
    \label{eq:QRT_AB}
\end{equation}
so that two-time correlators follow from the same Liouvillian $\mathcal{L}$.
In the stationary regime, $\rho(t)\to\rho_{\infty}$ and the correlator becomes invariant under time-translation,
\begin{equation}
    \langle A(t+\tau) B(t)\rangle
    = \langle A(\tau) B(0)\rangle
    = \Tr{%
        A\, e^{\mathcal{L}\tau}\!\left[ B \rho_{\infty} \right]}.
\end{equation}
The validity of Eq.~\eqref{eq:QRT_AB} relies on the same assumptions as the derivation
of the Markovian master equation (Born-Markov and secular approximation).
Beyond this regime, the QRT can fail~\cite{ford1996_no_qrt,Guarnieri2014,khan2022_qrt_multitime}.

Applying the QRT to the jump-current observable gives the regular part of the
two-time product $\langle I(t+\tau)I(t)\rangle$ for $\tau>0$ in terms of the same
Liouvillian that governs the average dynamics.
Inserting this result into the
definition of $G(\tau)$ and adding the equal-time shot-noise contribution of the
jump process gives
\begin{align}
    G(\tau)
    = \Tr{%
        \hat{I}\,
        e^{\mathcal{L}\tau}\!
        \left[\hat{I}\rho_\infty\right]
      }
      - \overline{I}^{\,2}
      + e\,\overline{I}\,\delta(\tau).
    \label{eq:G_of_tau_QPC}
\end{align}
The first term encodes dynamical relaxation governed by $e^{\mathcal{L}\tau}$,
the second subtracts the product of stationary means $\overline{I}^{\,2}$,
and the last term is the instantaneous counting shot noise from individual
tunneling events.

Measurable frequency-domain noise is obtained by transforming Eq.~\eqref{eq:G_of_tau_QPC} into the Fourier domain.
This transform is most transparent in the eigenbasis of the Liouvillian, where each decaying mode contributes a simple rational term weighted by its coupling to the current operator $\hat{I}$.
Using the spectral decomposition of the Liouvillian, with the right and left eigenvectors $|x_j\rangle\!\rangle$ and $|y_j\rangle\!\rangle$
satisfying $\mathcal{L}|x_j\rangle\!\rangle=\lambda_j|x_j\rangle\!\rangle$ and
$\langle\!\langle y_j|\mathcal{L}=\lambda_j\langle\!\langle y_j|$, normalized by
$\langle\!\langle y_i | x_j \rangle\!\rangle = \delta_{ij}$, and with the steady mode
$|x_0\rangle\!\rangle = |\rho_\infty\rangle\!\rangle$ ($\lambda_0=0$) and
$| y_0 \rangle\!\rangle = | \hone \rangle\!\rangle$, the noise spectrum can be written as~\cite{landi2024_current_fluctuations}
\begin{align}
    S(\w) = K - 2\Re \sum_{j\not=0} \frac{1}{\lambda_j-i\w}
    \langle\!\langle\hone|\hat{I}|x_j\rangle\!\rangle
    \langle\!\langle y_j | \hat{I} | \rho_\infty \rangle\!\rangle ,
    \label{eq:S(w)}
\end{align}
where
\begin{equation}
    K = e\,\overline{I}
    = e^2 \sum_{\alpha} \Tr{\hc_{\alpha}^\dagger \hc_{\alpha} \rho_\infty}.
\end{equation}
Then, the dimensionless Fano spectrum used for noise analysis in Sec.~\ref{sec:results} is defined as the noise spectrum normalized by the Poissonian value in the same counting-current convention,
\begin{equation}
    F(\omega) = \frac{S(\omega)}{e \, \overline{I}}.
    \label{eq:Fano_spectrum}
\end{equation}
The zero-frequency value $F\equiv F(0)$ is the Fano factor, defined as the ratio of zero-frequency noise to the Poissonian noise value.
For purely Poissonian transport, one has $F=1$, while $F<1$ ($F>1$) signals
sub-Poissonian (super-Poissonian) statistics and indicates correlations (anti-correlations) between
successive tunneling events~\cite{blanter2000_shot_noise,clerk2010_rmp_noise,landi2024_current_fluctuations}.
In the present context, the average current fixes the stationary detector response, while the full frequency dependence of $F(\w)$ reflects the relaxation mechanisms that modulate the QPC current within the singlet or triplet steady-state sector.
Therefore the steady-state Fano spectra serve as dynamical fingerprints of the microscopic pathways that remain visible in the measured current.

Equation~\eqref{eq:S(w)} also gives the diagnostic used below to identify which microscopic paths dominate the low-frequency noise.
Each non-stationary Liouvillian mode contributes with a denominator $\lambda_j-i\w$ and with a current weight
\begin{equation}
    W_j =
    \langle\!\langle\hone|\hat{I}|x_j\rangle\!\rangle
    \langle\!\langle y_j | \hat{I} | \rho_\infty \rangle\!\rangle.
\end{equation}
Thus, a path can dominate the noise only if it feeds a slow mode, with small $-\Re\lambda_j$, and if that mode is visible in the measured QPC current, i.e., if $W_j$ is not negligible.
For this reason, the dominant paths cannot be inferred from the largest jump rates alone.
In post-processing, the reduced-sector Liouvillian is diagonalized first, and the slow population-carrying modes are retained.
For a transition pair $a\leftarrow b$, all counted-current jump channels that transfer the population from the reduced eigenstate $b$ to $a$ are grouped into a superoperator $\mathcal{J}_{a\leftarrow b}$.
The importance of this pair for the mode $j$ is ranked by the product of its excitation of the slow mode from the steady state and the current visibility of that mode,
\begin{equation}
    \mathcal{S}^{(j)}_{a\leftarrow b}
    =
    \left|
    \langle\!\langle y_j |
    \mathcal{J}_{a\leftarrow b}
    | \rho_\infty\rangle\!\rangle
    \right| \times
    \left|
    \langle\!\langle\hone|\hat{I}|x_j\rangle\!\rangle
    \right|.
    \label{eq:rank of the mode}
\end{equation}
This score is used only as a controlled ranking, not as a unique decomposition of $S(\w)$ into individual arrows.
The ranked pairs are then checked against two additional quantities.
The first is the effective dynamical population-transfer rate generated by the dissipative jump operators,
\begin{equation}
    \Gamma_{a\leftarrow b}
    =
    \sum_{\alpha}
    \left|
    \mel*{a}{\hc_\alpha}{b}
    \right|^2,
\end{equation}
where the sum runs over the jump operators retained in the steady-state sector under consideration, singlet or triplet [cf.~Fig.~\ref{fig:jumps-by-sector}].
The second is the current contrast between the two reduced eigenstates,
\begin{equation}
    \begin{aligned}
    \Delta I_{a b} &= I_a-I_b,\\
    I_a &= \Tr{\hat{I}\dyad*{a}{a}}
    =
    e\sum_{\alpha}
    \mel*{a}{\hc_\alpha^\dagger\hc_\alpha}{a},
    \end{aligned}
\end{equation}
with the same sector-restricted sum over jump operators.
Thus $\Delta I_{a b}$ measures how strongly the detector current distinguishes the two states connected by the transition pair; it is a visibility criterion rather than a transition rate.
The pathways quoted in the results [Sec.~\ref{sec:results}] are precisely the pairs that occur in slow population modes, have appreciable $\mathcal{S}^{(j)}_{a\leftarrow b}$, and connect states with a visible current contrast.
Finally, the reduced connected pathway models in Fig.~\ref{fig:f0-vs-eps} provide a numerical validation of this interpretation: when only the selected pathways, weak connector channels, and elastic background are retained, the resulting $F(0)$ closely follows the full model.

\begin{figure*}[t!]
\begin{minipage}{0.47\linewidth}
    \centering
    \includegraphics[width=\linewidth]{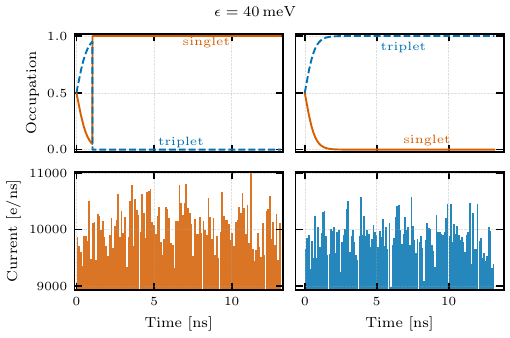}
\end{minipage}
\hfill
\begin{minipage}{0.47\linewidth}
    \centering
    \includegraphics[width=\linewidth]{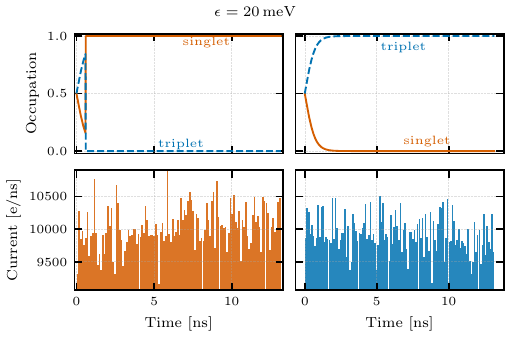}
\end{minipage}
\caption{
  Representative quantum-jump trajectories for post-selected singlet and triplet branches.
  Left: $\eps=40$~meV, i.e., above the QPC bias.
  Right: $\eps=20$~meV, i.e., below the QPC bias.
  Each panel shows the corresponding occupation dynamics together with the detector current record for a single representative trajectory in the singlet and triplet branches.
  The plotted QPC current is shown as a time histogram --- counted QPC events are summed in 120 equal-width bins and divided by the bin width, corresponding to $\Delta t\simeq0.11$~ns per displayed current bin.}\label{fig:time-evolution-branches}
\end{figure*}

\section{Results}\label{sec:results}

\subsection*{Time evolution}

First, consider the time-domain detector signal generated from the coherent initial state
\begin{align}
    \ket*{\psi_0} = \frac{1}{\sqrt{2}}\bigl( \ket*{S_0^g} + \ket*{T^g} \bigr) = \ket*{L^g R^g}
\end{align}
and post-select trajectories according to the singlet-like or triplet-like steady state that they approach at long times.
This choice of initial condition is natural for readout because it corresponds to a coherent superposition of the two branches that the detector is meant to distinguish.
For each trajectory, both the state occupations and the full stochastic current record of the QPC are monitored.

Unless stated otherwise, the on-site Coulomb energy is fixed at $U=1$~meV, the QPC bias at $V=30$~meV, the interdot tunneling amplitude at $t=0.05$~meV, and the QPC tunneling parameters at $\mathcal{T}=\sqrt{0.3}\approx0.548$ and $\nu=-0.15\,\mathcal{T}\approx-0.082$.
Two representative level spacings are considered, $\eps=40$~meV and $\eps=20$~meV, corresponding to the regimes above and below the QPC bias window, respectively.
The steady-state current spectra in Fig.~\ref{fig:steady-fano-comparison} are obtained by averaging over 2000 quantum-jump trajectories for each post-selected branch.
Each trajectory is evolved up to $t_{\max}/\hbar = 3.0\times10^7/U$, and the numerical Fourier transform is applied only to the final half of the record, after discarding the initial transient.

Figure~\ref{fig:time-evolution-branches} shows representative quantum-jump trajectories for two level spacings chosen on opposite sides of the detector bias window.
For each value of $\eps$, the figure displays one representative stochastic realization for the post-selected singlet branch and one for the post-selected triplet branch.
Each panel contains the corresponding occupation traces together with the time-dependent QPC current recorded along the same trajectory, so the plotted jumps show how changes in the DQD state are accompanied by changes in the detector signal.
These examples are included only to illustrate the structure of the post-selected trajectories in the time domain.
The current statistics and noise spectra are analyzed quantitatively in Fig.~\ref{fig:steady-fano-comparison} and discussed below.

\subsection*{Steady state analysis}

\begin{figure}[t!]
    \centering
    \includegraphics[width=\linewidth]{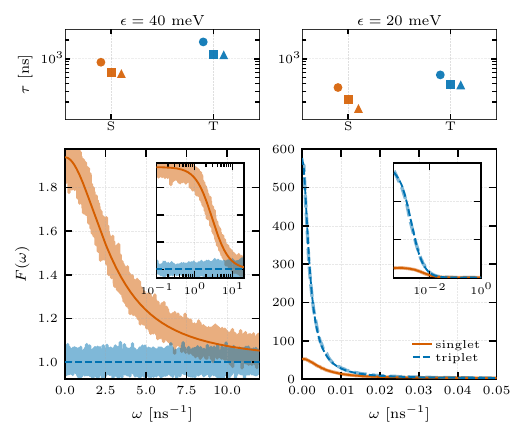}
    \caption{Steady-state current Fano spectra obtained from the Liouvillian formalism and from long-time quantum-jump trajectories.
    The left column corresponds to $\eps=40$~meV and the right column to $\eps=20$~meV.
    The upper panels summarize the relaxation times $\tau$ of the three slowest Liouvillian modes for the two branches in each case.
    The lower panels show the Fano spectra on a linear frequency axis, while the insets display the same spectra on a logarithmic frequency axis to resolve the low-$\w$ structure.
    The singlet spectrum (solid vermilion line) and the triplet spectrum (dashed blue line) are compared with the corresponding numerical trajectory estimates.}\label{fig:steady-fano-comparison}
\end{figure}

To understand fluctuations in a stationary setting, the QPC Fano spectrum $F(\w)$ is computed in two complementary ways.
First, the Liouvillian expression for $F(\w)$ is evaluated in the reduced singlet and triplet sectors and normalized according to Eq.~\eqref{eq:Fano_spectrum}.
Second, long quantum-jump ensembles are generated, and the same normalized quantity is extracted numerically from the trajectory records.
Figure~\ref{fig:steady-fano-comparison} overlays these two approaches and therefore serves as the main consistency check between steady-state theory and direct stochastic simulations.
The plotted quantity is the steady-state Fano spectrum, shown separately for the singlet and triplet branches.

The microscopic assignments in the following paragraphs use the modal pathway analysis described in Sec.~\ref{sec:methods}.
The relaxation-time panels in Fig.~\ref{fig:steady-fano-comparison} identify which Liouvillian modes are slow enough to generate low-frequency noise, but they do not by themselves identify the responsible jumps.
For each slow mode $|x_j\rangle\!\rangle$, the dominant reduced-basis populations are inspected, the transition pairs are ranked by $\mathcal{S}^{(j)}_{a\leftarrow b}$ [Eq.~\eqref{eq:rank of the mode}], and only those pairs that also have a visible QPC-current contrast are retained.
This is why some fast rearrangements within closely connected pairs of states appear prominently in the jump-rate matrices of Fig.~\ref{fig:jumps-by-sector} but are not interpreted as the source of the low-frequency excess --- they equilibrate local state pairs, while the noise is generated by slower current-contrasting transfers between shelved and conducting parts of the network.

The lower-left panel of Fig.~\ref{fig:steady-fano-comparison} shows that for $\eps=40$~meV the triplet represents the conventional PSB limit, that is, its Fano spectrum is flat and stays close to the Poissonian value $F(\w)=1$ throughout the resolved frequency range.
The singlet behaves differently.
The singlet spectrum instead retains a broad low-frequency enhancement, showing that slow internal dynamics within the singlet manifold remains visible in the current fluctuations even after the mean current has reached its steady value.
The upper-left panel should therefore not be read as an absence of slow triplet eigenmodes (the plotted triplet relaxation times are in fact longer than the singlet ones).
The flat triplet spectrum instead shows that, at $\eps=40$~meV, these triplet modes have negligible current visibility, whereas the singlet slow modes remain visible in the connected current correlator.
Microscopically, the residual singlet excess noise comes from weak but current-contrasting intra-manifold shelf cycles, most visibly the ground-manifold chain $s_5 \leftrightarrow s_2 \leftrightarrow s_6$ and the excited-manifold chain $s_9 \leftrightarrow s_4 \leftrightarrow s_{10}$.
Thus, above the bias threshold the excited level leaves only a weak imprint on the triplet sector, while the singlet retains a broadened low-frequency response, consistent with the earlier QPC readout analyses of Refs.~\cite{Stace2004,Barrett_PRB_2006,StaceBarretArxiv,marcinowski_phonon_2013,roszak_decoherence-enhanced_2015}.

The lower-right panel reveals the more interesting low-$\eps$ regime.
Here, the low-frequency noise becomes strongly enhanced in both sectors, but the triplet develops the sharper and larger zero-frequency feature.
The upper-right panel shows the relevant time-scale contrast.
That is, the triplet is controlled by one dominant slow mode, whereas the singlet involves several comparable slow modes.
Physically, the triplet branch is governed by a shelving mechanism.
Rare transitions involving $t_4$, mainly $t_4 \leftrightarrow t_2$ and $t_4 \leftrightarrow t_3$, connect the low-current shelved state to a more conducting part of the triplet network; within that active sector, the fast process $t_2 \leftrightarrow t_5$ mainly equilibrates the current-carrying states.
Because entry into and escape from that conducting branch are rare, the detector current switches telegraphically and the corresponding noise accumulates very close to $\w=0$.
The logarithmic-frequency insets in Fig.~\ref{fig:steady-fano-comparison} make this low-$\w$ structure particularly visible.
The singlet remains super-Poissonian as well, but its excess noise is broader and less sharply concentrated near $\w=0$, reflecting a different pattern of internal relaxation.
In the singlet branch, by contrast, the current is modulated by several current-contrasting intra-manifold shelf cycles, namely $s_5 \leftrightarrow s_2 \leftrightarrow s_6$, $s_7 \leftrightarrow s_1 \leftrightarrow s_8$, and $s_9 \leftrightarrow s_4 \leftrightarrow s_{10}$, without a single comparably long-lived shelved state dominating the dynamics.
The detector therefore samples a broader distribution of relaxation times, which spreads the excess noise over a wider frequency window instead of concentrating it into one narrow low-frequency peak.
This distinction is important.
The two branches are not merely ``more'' or ``less'' noisy, but encode different dynamical mechanisms in the shape of their spectra.

The quantum-jump estimates reproduce the Liouvillian spectra well over the resolved frequency window.
The remaining deviations are restricted to the extremely narrow structure closest to $\w=0$, where the correlation times become so long that resolving the exact limit demands very long trajectories.
In that sense, the comparison is already informative even before perfect convergence at $\w=0$ is achieved.
The broad spectral shape and the hierarchy between singlet and triplet branches are robust, while the unresolved residual mismatch is consistent with the presence of ultra-slow dynamics.

\begin{figure}[t!]
    \centering
    \includegraphics[width=\linewidth]{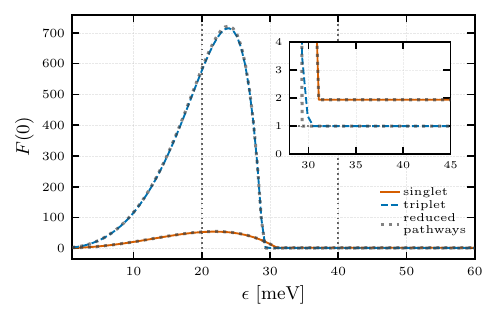}
    \caption{
      Zero-frequency Fano factor $F(0)$ as a function of the level spacing $\eps$.
      The singlet branch (solid vermilion line) and the triplet branch (dashed blue line) are shown together.
      The dark-gray dotted curves show reduced connected pathway models constructed from the dominant shelving transitions and weak connector channels selected by the modal pathway ranking of Sec.~\ref{sec:methods}.
      Both reduced models also retain the elastic background channel needed to close the dynamics.
      The inset highlights the range where the singlet remains super-Poissonian while the triplet has already relaxed back to Poissonian behavior.}
    \label{fig:f0-vs-eps}
\end{figure}

Finally, Fig.~\ref{fig:f0-vs-eps} condenses this behavior into the zero-frequency Fano factor as a function of the level spacing.
The figure shows a clear crossover as $\eps$ is reduced.
For large $\eps$, the triplet branch rapidly approaches the featureless Poissonian QPC background associated with an effectively blocked DQD branch, whereas the singlet branch remains super-Poissonian.
Decreasing $\eps$ gradually activates excited-level-assisted pathways, and the triplet noise increases sharply until it overtakes the singlet.
The additional dark-gray dotted curves show reduced connected pathway models built from the dominant shelving transitions identified by the Liouvillian analysis, together with the weak connector channels and elastic background needed to close the network dynamically.
For the singlet branch, the reduced pathway model retains the broad enhancement almost unchanged, confirming that the low-frequency singlet noise is likewise a shelving effect, but distributed over several connected shelf cycles rather than concentrated in a single dominant switch.
For the triplet branch, the dark-gray dotted reduced curve lies almost on top of the full result as well, showing that the giant low-frequency excess is already captured by one minimal shelving network built around the $t_4$ bottleneck and the conducting $t_2/t_5$ branch.
The inset highlights an intermediate regime in which the singlet is still clearly super-Poissonian while the triplet has already decayed toward $F(0)\simeq1$.
Taken together, Figs.~\ref{fig:time-evolution-branches}--\ref{fig:f0-vs-eps} show that the level spacing controls not only the magnitude of the current noise but also which branch is noisier and on what time scale the detector fluctuations are generated.

\section{Conclusions}\label{sec:conclusions}

Extending the standard single-level picture of Pauli spin-blockade readout to a DQD with ground and excited single-particle levels does not merely renormalize the detector response, but can qualitatively change its noise statistics.
When the relevant excited-level transitions enter the QPC bias window, the triplet is no longer an ideally blocked configuration.
Detector-induced transitions involving mixed and excited configurations open additional pathways, so the current fluctuations acquire a slow-switching component on top of the Poissonian shot-noise background.
In that regime, the low-frequency part of the spectrum carries the clearest signature of the underlying level structure.

The key control parameter is the position of the level spacing $\eps$ relative to the QPC bias window.
When the excited-level-assisted transitions are energetically suppressed, the familiar single-level PSB phenomenology is largely recovered.
The triplet branch stays close to Poissonian detector statistics, while the singlet remains super-Poissonian because charge rearrangement within the singlet manifold still modulates the detector current.
However, for smaller $\eps$, the QPC can activate rare transitions through additional excited-state configurations.
These rare switching events act as a shelving mechanism that generates long correlation times and strong enhancement of the zero-frequency noise.
As a result, the triplet branch can become noisier than the singlet, even though single-level intuition would suggest the opposite ordering.

The comparison between long quantum-jump trajectories and the Liouvillian calculation shows that this interpretation is consistent across both approaches.
The trajectory data reproduce the resolved frequency dependence of the Fano spectra over the accessible frequency window, while the Liouvillian treatment makes explicit that the giant low-frequency enhancements are tied to very slow relaxation modes.
In this sense, the QPC noise is not only a readout metric but also a diagnostic of hidden leakage channels and their characteristic time scales.

From a device-design perspective, these results identify the level spacing as a central parameter for reliable PSB readout.
If one aims to preserve the conventional PSB readout contrast, the detector bias should remain below the onset of excited-level-assisted transitions, or the corresponding excited-level admixture should be suppressed by design.
In contrast, operating close to that threshold provides a sensitive way to diagnose excited-level leakage through the emergence of pronounced low-frequency super-Poissonian noise.
This makes current-noise spectroscopy a useful complementary tool to average-current measurements in spin-qubit devices with accessible excited levels.

\begin{acknowledgments}
The author thanks Dr.\, Katarzyna Roszak for insightful
discussions and for sharing preparatory notes that helped frame this
study.

This work was supported by the Marie Sk\l{}odowska-Curie Actions COFUND project, co-funded by the European Union (Physics for Future --- Grant Agreement No.~101081515).
\end{acknowledgments}

\bibliography{references}

\end{document}